\documentclass[aps,nofootinbib,floatfix,showpacs,preprintnumbers,twocolumn]{revtex4} %eqsecnum,

\def\be{\begin{equation}}
\def\ee{\end{equation}}
\def\bea{\begin{eqnarray}}
\def\eea{\end{eqnarray}}
\newcommand{\ec}[1]{Eq.~(\ref{eq:#1})}
\newcommand{\Ec}[1]{(\ref{eq:#1})}

\newcommand{\eql}[1]{\label{eq:#1}}

\newcommand{\vs}{\nonumber\\}

\def\square{\kern1pt\vbox{\hrule height 1.2pt\hbox{\vrule width 1.2pt\hskip 3pt
   \vbox{\vskip 6pt}\hskip 3pt\vrule width 0.6pt}\hrule height 0.6pt}\kern1pt}
   
\usepackage{graphicx}

\newcommand{\rf}[1]{\ref{fig:#1}}
\newcommand\osq{\overline{\square}}
\newcommand\ovr{\overline{R}}

\begin{document}

%%%%%%%%%%%%%%%%%%%%%%%%The update log%%%%%%%%%%%%%%%%%%

%\begin{flushright}
%Feb. 25. 2012, spark 
%May 29, 2012, spark
%Aug. 2, 2012, spark
%Sep. 3, 2012, spark updated from v3
%Dec. 18, 2012, spark revised from v4 for resubmission
%\end{flushright}

%%%%%%%%%%%%%%%%%%%%%%%%%%%%%%%%%%%%%%%%%%%%%%%%%%%%%%%%%%%%%
\title{Structure formation in a nonlocally modified gravity model}         % Enter your title between curly braces
\author{Sohyun Park$^{1, 2}$ and Scott Dodelson$^{2, 3, 4}$  }        % Enter your name between curly braces
%\date{}          % Enter your date or \today between curly braces
\affiliation{${}^1$Department of Physics, University of Florida, Gainesville, FL 32611}
\affiliation{${}^2$Center for Particle Astrophysics, Fermi National Accelerator Laboratory, Batavia, IL 60510}
\affiliation{${}^3$Department of Astronomy and Astrophysics, University of Chicago,
Chicago, IL 60637}
\affiliation{${}^4$Kavli Institute for Cosmological Physics, University of Chicago, Chicago, IL 60637}

\begin{abstract}
We study a nonlocally modified gravity model proposed by Deser and Woodard\cite{Deser:2007jk} which gives an explanation for current cosmic acceleration. By deriving and solving the equations governing the evolution of the structure in the Universe, we show that this model predicts a pattern of growth that differs from standard general relativity ($+$dark energy) at the 10-30\% level. These differences will be easily probed by the next generation of galaxy surveys, so the model should be tested shortly.
\end{abstract}
\maketitle

%%%%%%%%%%%%%%%%%%%%%%%The End of Title Page%%%%%%%%%%%%%%%%%%%%%%%%%%%%

\section{Introduction}

There are multiple threads of evidence -- from distant supernovae~\cite{Riess:1998,Perlmutter:1999}, the cosmic microwave background~\cite{Dunkley:2009,Komatsu:2009,Sherwin:2011gv,vanEngelen:2012va}, the galaxy distribution~\cite{Sanchez:2012sg}, and correlations of these with each other~\cite{Scranton:2003in} -- that the Universe is accelerating 
The multiple sets of evidence argue against any one systematic affecting the qualitative conclusion that $\ddot a>0$ where $a$ is the scale factor of the Universe. 

General relativity maintains intact our Newtonian intuition that gravity should lead to deceleration, in the case of the smooth Universe that $\ddot a$ should be negative. The one caveat is that general relativity allows not only energy, but also pressure, to impact geometry, and a substance with negative pressure can drive acceleration. One possibility then is that the multiple strands of cosmological evidence point to the need for a new, previously undiscovered substance with negative pressure, so-called {\it dark energy}~\cite{Frieman:1995pm,Coble:1996te,Caldwell:1997ii,Ratra:1998,Wetterich:1998,Huterer:1998qv}. 

Another possibility is that general relativity is wrong~\cite{Carroll:2003wy}. The theory was developed using information and intuition honed on relatively small scales (the Solar System), so it would not be shocking if it needs to be modified on the larger scales probed by cosmology. There have several problems so far with this approach. Perhaps, most generally, we do not have a set of simple guidelines, thought experiments, and arguments of elegance that motivated Einstein as he constructed general relativity. Even if beauty is abandoned, a modification of gravity must still confront three problems, two of which are related. First, almost all models contain a mass scale that must be set to be much smaller than any mass found in nature, less than $10^{-33}$ eV. Why such a small mass scale exists, and how it can be protected in the presence of even weak interactions with the rest of physics, is a profound problem. Related to this is the fine tuning problem in time that afflicts these models: the modifications to gravity just happen to be important today, not at any time in the past. Believing that we live in a special time is yet another price we pay for these modifications. A third problem with modified gravity models is that they must be introduced in a way that does not violate the successes of general relativity in the Solar System. Indeed, these constraints~\cite{Chiba:2003ir} doomed one of the first modified gravity models introduced to explain acceleration and still place tight constraints on many others
(e.g., see~\cite{Nojiri:2007,Saini:2007,Guo:2005}). 

One class of modified gravity models that rises above these challenges contains non-local interactions\cite{Deser:2007jk}, i.e., terms in the Lagrangian that depend on the values of the fields at more than one point in space-time. In particular, Deser and Woodard~\cite{Deser:2007jk} pointed out the utility of terms that are functionals of $\square^{-1} R$, where $\square$ is the d'Alembertian and $R$ the Ricci scalar. 
%%%Since $R$ is zero in the radiation dominated era, these terms naturally are irrelevant at early times and begin to affect the dynamics only after the matter-radiation transition, 
%%%%%%%%%%%%%%%%%%%%%%%%%%%%%%%%%%%%%%%%%%%%%%%%%%%%%%%%%%%%%%%%%%%%%%%%%%%
In the cosmological context, $\square^{-1} R$ grows very slowly: as $(t/t_{eq})^{1/2}$ in the radiation dominated era and logarithmically in the matter dominated era. So, at the time of Nucleosynthesis $\square^{-1} R$ is about $10^{-6}$ and at matter-radiation
equality it is only order one.  These terms therefore naturally are irrelevant at early times and begin to affect the dynamics of the Universe only after the matter-radiation transition, 
%%%%%%%%%%%%%%%%%%%%%%%%%%%%%%%%%%%%%%%%%%%%%%%%%%%%%%%%%%%%%%%%%%%%%%%%%%%
thereby circumventing some of the fine tuning difficulties. Since $\square^{-1} R$ is dimensionless, the functional that multiplies $R$ has no new mass parameter. Finally, because $\square^{-1}R$ is extremely small in the Solar System, these models easily pass local tests of gravity. 

There is some theoretical motivation for these kinds of terms. Away from the critical dimension in string theory, $R\square^{-1}R$ is precisely the term generated by quantum corrections, as first pointed out by Polyakov\cite{Polyakov:1981rd}. More generally, there have been several arguments for considering non-local theories~\cite{Parker:1985kc,Banks:1988je,Hamber:2005dw,Biswas:2006,Barnaby:2010kx}. 
It might be possible to rewrite these models in terms of local models with an auxiliary scalar field~\cite{Capozziello:2008gu,Koshelev:2008ie}.  Given that they can explain the current epoch of acceleration, it is not surprising that an early epoch of inflation has also been attributed to non-local interactions~\cite{Tsamis:1997rk,Nojiri:2007uq,Prokopec:2011ce}. A realistic model for acceleration was constructed in Ref.~\cite{Deffayet:2009ca}, where they chose the arbitrary function of $\square^{-1}R$ so that it reproduced the exact expansion history generated by standard $\Lambda$CDM. This working model demonstrated the feasibility of the non-local approach to generating acceleration and therefore we focus on it here, but of course it also points to a problem inherent in this way (and many others) of modifying gravity: there is a completely arbitrary free function that can be chosen at will to fit the expansion history.

One of the most intriguing aspects of modified gravity models in general is that, even if they are constructed to reproduce the expansion history equivalent to that of a given dark energy model (such as $\Lambda$CDM), perturbations will often evolve differently than in a model with standard general relativity and dark energy. This has led to the idea that the way to distinguish dark energy models from modified gravity models is to measure the growth of structure in the Universe. While there has been some work to date on perturbations in non-local models~\cite{Koivisto:2008dh}, there have been no concrete predictions for the growth of structure and how this may differ from dark energy models. Here we work out the perturbation equations in this new class of models and solve perturbatively to arrive at a concrete prediction for two observables as a function of redshift. The deviations from dark energy models are at the 10-30\% level and have a characteristic signature as a function of redshift, which suggests that the class of models will be tested by upcoming surveys.

The rest of the paper is organized as follows: \S\ref{sec2} reviews the model of Ref.~\cite{Deffayet:2009ca} and presents the equations governing the expansion history in this model. The next section perturbs the model around this zero order solution to arrive at the equations that govern the growth of structure. The next two sections solve these equations perturbatively for two quantities: $\Psi + \Phi$ and $G_{\rm eff}$ where $\Psi$ and $\Phi$ are the two scalar potentials in the perturbed Friedman Robertson Walker metric and $G_{\rm eff}$ is the proportionality constant between $\nabla^2\Phi$ and $4\pi\rho\delta$ in the Poisson equation. In general relativity, $\Psi+\Phi=0$ and $G_{\rm eff}=G_N$, Newton's constant, but in modified gravity models these equalities often do not hold. Indeed, we show in \S\ref{sec4} and \S\ref{sec5} that the deviations in this non-local model are significant and measurable.

\section{Model and Zeroth Order Equation}\label{sec2}

\begin{figure}[tbh]
\includegraphics[width=7.5cm]{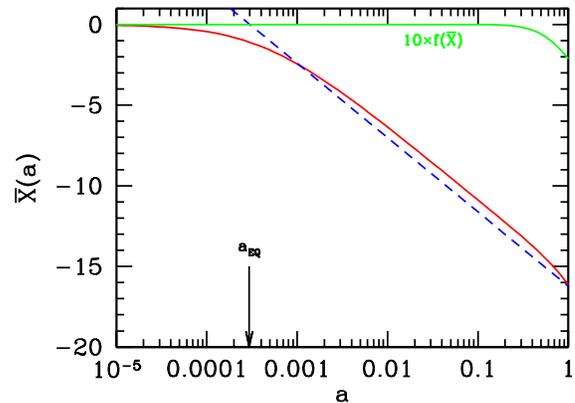}
\caption{The evolution of $\overline{X} \equiv \osq^{-1}\overline{R}$. Dashed line shows the analytic solution in the matter era: $\overline{X} = -2\ln(a/a_{\rm EQ})$. The flat (green) curve shows the function $f(\bar X)$ chosen in Ref.~\cite{Deffayet:2009ca} to fit the $\Lambda$CDM expansion history. Its amplitude and shape are chosen so that it has negligible impact on the expansion until relatively recently.}
\label{barx}
\end{figure}

We take a model in which the Einstein-Hilbert Lagrangian is nonlocally modified as \cite{Deser:2007jk}
\begin{equation}
 \mathcal{L}_g = \frac{1}{16\pi G}\sqrt{-g}R\bigg[1 + f(\frac{1}{\square}R)\bigg]
 \eql{action}
\end{equation}
where $f$ is a function of its dimensionless argument that will be determined by matching the expansion history of $\Lambda$CDM. To understand how this new term helps drive acceleration today, note that in a smooth expanding Universe described by the Friedmann-Robertson-Walker metric, the non-local operator at time $t$ is a double integral over all previous times:
\bea
\overline{X} &\equiv& \osq^{-1}\overline R =-\int_{t_i}^{t}\frac{d t'}{a^3(t')}\int_{t'_i}^{t'}d t'' a^3(t'') \overline{R}(t'')
\vs
&=& -\int_{t_i}^{t}\frac{d t'}{a^3(t')}\int_{t_i}^{t'}d t'' a^3(t'')
\Bigl[ 6\dot{H}(t'') + 12H^2(t'') \Bigr] \vs
\eql{defx}
\eea
% R = 6(\ddot a/a + H^2) = 6( \dot H + 2H^2)
% \dot H = \ddot a/a - \dot a^2/a^2 = \ddot a/a - H^2
where here and throughout we denote quantities in the smooth FRW background with over-bars, e.g. $\overline{R}$, to distinguish them from the full perturbed quantities; and $H\equiv \dot a/a$ is the expansion rate. 
%%%In a radiation dominated Universe, $\ovr=0$. 
%%%%%%%%%%%%%%%%%%%%%%%%%%%%%%%%%%%%%%%%%%%%%%%%%%%%%%
In a perfect radiation dominated Universe, $\ovr=0$.
%%%%%%%%%%%%%%%%%%%%%%%%%%%%%%%%%%%%%%%%%%%%%%%%%%%%%%%
So, in the standard cosmological picture, with radiation giving way to matter at $a_{\rm EQ} = 3\times 10^{-4}$, $\overline{X}\equiv\osq^{-1}\ovr$ is very small until $a_{\rm EQ}$ after which it grows logarithmically so remains of order unity today (solid curve in Fig.~\rf{barx}). During the epoch in which the Universe is purely matter dominated, $a\propto t^{2/3}$ so  $\osq^{-1}\ovr\big\vert_{\rm MD} = -2\ln(a/a_{\rm EQ})$, an approximation that Fig.~\ref{barx} shows captures the growth at late times.

%\Sfig{barx}{The evolution of $\overline{X} \equiv \osq^{-1}\overline{R}$. Dashed line shows the analytic solution in the matter era: $\overline{X} = -2\ln(a/a_{\rm EQ})$. The flat (green) curve shows the function $f(\bar X)$ chosen in Ref.~\cite{Deffayet:2009ca} to fit the $\Lambda$CDM expansion history. Its amplitude and shape are chosen so that it has negligible impact on the expansion until relatively recently.}

It is straightforward to choose a function of $X\equiv \square^{-1}R$ in the action of \ec{action} that begins to modify the expansion history only at late times when its argument becomes of order unity. To see what form the function must take, we vary the action with respect to the metric, obtaining the modified Einstein equations:
\begin{equation}
G_{\mu\nu} + \Delta G_{\mu\nu} = 8\pi G T_{\mu\nu} \;.
\label{nonlocal_field_eq}
\end{equation}
Here the correction to the Einstein tensor, $\Delta G_{\mu\nu}$ is~\cite{Deser:2007jk}
\begin{eqnarray}
\Delta G_{\mu\nu} &=& \Bigl[ G_{\mu\nu} + g_{\mu\nu}\square - D_{\mu}D_{\nu} \Bigr]
\Biggl\{ f(X)  +\frac{1}{\square}\Bigl[R f'(X)\Bigr]\Biggr\}
\vs
&+&\Bigl[ \delta_{\mu}^{(\rho}\delta_{\nu}^{\sigma)} 
- \frac{1}{2}g_{\mu\nu}g^{\rho\sigma}\Bigl] 
\partial_{\rho}\Bigr(X\Bigl)
\partial_{\sigma}\bigg(\frac{1}{\square}\Bigl[R f'\Bigr(X\Bigl)\Bigr]\bigg)\;.
\vs
\eql{DeltaGmn}
\end{eqnarray}
Applying these general equations to the FRW metric leads to
\begin{eqnarray}
3H^2 + \Delta G_{00} &=& 8\pi G \rho \\
-2\dot{H} -3H^2 + \frac{1}{3a^2}\delta^{ij}\Delta G_{ij} &=& 8\pi G P .
\end{eqnarray}
Here $\rho$ and $P$ are the energy density and pressure and the zero order (i.e, homogeneous) nonlocal terms are \cite{Deser:2007jk}
\begin{eqnarray}
\Delta G_{00} &\!\!\!=\!\!\!& 
\Bigl[ 3H^2 + 3H\partial_t]
\Biggl\{ f \Bigr(\overline{X}\Bigl) + \frac{1}{\overline{\square}}\Bigl[\overline{R} f'\Bigr(\overline{X}\Bigl)\Bigr]\Biggr\}
\vs
&&+\frac{1}{2}
\partial_{t}\Bigr(\overline{X}\Bigl)
\partial_{t}\bigg(\frac{1}{\overline{\square}}\Bigl[\overline{R} f'\Bigr(\overline{X}\Bigl)\Bigr]\bigg)\;, \\
\Delta G_{ij} &\!\!\!=\!\!\!& 
+a^2\delta_{ij}\Bigg[ 
\frac{1}{2}
\partial_{t}\Bigr(\overline{X}\Bigl)
\partial_{t}\bigg(\frac{1}{\overline{\square}}\Bigl[\overline{R} f'\Bigr(\overline{X}\Bigl)\Bigr]\bigg)\;
\vs
&\!\!\!-\!\!\!&\Big( 2\dot{H}+ 3H^2+2H\partial_t + \partial^2_t\Big)
\left( f %\Bigr(\overline{X}\Bigl)
+ \frac{1}{\overline{\square}}\Bigl[\overline{R} f'\Bigr]\right)
\Bigg]
.\vs
\label{DeltaGmn_zeroth}
\end{eqnarray}
Carrying out the reconstruction program using this revised Friedmann equation, Deffayet and Woodard~\cite{Deffayet:2009ca} showed that the function
\be
f(X) = 0.245\Bigl[\tanh(0.350Y+0.032Y^2 +0.003Y^3) -1 \Bigr] \;,
\eql{deff}
\end{equation}
where $Y \equiv X + 16.5$ leads to an expansion history equivalent to that produced by $\Lambda$CDM when the rest of the cosmological parameters are set to their default values, 
$\{\Omega_{\rm curv}, \Omega_{m}, \Omega_{r}\} = \{0, 0.28, 8.5 \times 10^{-5}\}$. 
%%%%%%%%%%%%%%%%%%%%%%%%%%%%%%%%%%%%%%%%%%%%%%%%%%%%%%%%%%%%%%%%%%%%%%%%%%%%%%%%%%%
%$\{\Omega_{\Lambda}, \Omega_{m}, \Omega_{r}\} = \{0.72, 0.28, 8.5 \times 10^{-5}\}$.
%%%%%%%%%%%%%%%%%%%%%%%%%%%%%%%%%%%%%%%%%%%%%%%%%%%%%%%%%%%%%%%%%%%%%%%%%%%%%%%%%%%
This function, called the {\it non-local distortion function}, is depicted in Fig.~\ref{barx}, which gives a sense of the fine-tuning needed in this model. The function must be very flat so that the new terms do not alter the expansion history until only recently. The time at which the modifications kick in and the amplitude are both fine-tuned to produce the correct expansion history. Due to the logarithmic nature of the growth of $\osq^{-1}\ovr$, though, this fine-tuning is many orders of magnitude less severe than that in most other models. Apart from aesthetic arguments, the model defined by this function is concrete enough to enable predictions to be made for the growth of structure. 

\section{Perturbation Equations}

In order to follow the growth of inhomogeneities in the universe, we write the metric as the background FRW geometry plus small perturbations. Plugging the metric into the field equation (\ref{nonlocal_field_eq}) and expanding it to the first order gives the evolution equations for the perturbations. We consider only scalar perturbations and follow the notation of Ref.~\cite{Dodelson} (Eq. (4.9) there):
\begin{eqnarray}
g_{00}(t, \vec{x}) &=& -1 -2\Psi(t,\vec x) \;, \quad
g_{0i}(t, \vec{x}) = 0 \;, \vs
g_{ij}(t, \vec{x}) &=& \delta_{ij} a^2(t) 
\Bigr[1 + 2\Phi(t,\vec x)\Bigl]\;.
\end{eqnarray}
With these ingredients the modified field equations at first order are
\begin{eqnarray}
\delta\Bigl( G_{00} + \Delta G_{00}\Bigr) &=& 
8\pi G  \delta  {T}_{00}\eql{00_delta_eq}\\
\delta\Bigl( G_{ij} + \Delta G_{ij}\Bigr) &=& 
8\pi G \delta  {T}_{ij}\
\eql{ij_delta_eq}.
\end{eqnarray}
In our notation, little $\delta$ refers to first order perturbations, and capital $\Delta$ means the correction to the original Einstein equation due to the non-local terms. The time-time component leads to the modified Poisson equation on small scales, and we focus on that modification first. The space-space components can be arranged to yield an equation relating the two scalar potentials, and we treat the non-local modification of that equation afterwards. 

\subsection{Modified Poisson Equation}

Some of the terms in the time-time component of \ec{00_delta_eq} are standard; for example, in Fourier space, 
\begin{eqnarray}
\delta G_{00} &\!\!\!=\!\!\!& \Bigl[6H\dot{\Phi} + 2\frac{k^2}{a^2}\Phi\Bigr] \rightarrow 2\frac{k^2}{a^2} \Phi\;\vs
\delta T_{00} &=& \bar\rho\delta.
\end{eqnarray}
On the first line, we drop the first term because we are working in the sub-horizon limit so the second term, which is of order $(k/H)^2$ with respect to the first, dominates. On the second line, $\bar\rho$ is the mean density of matter, and $\delta$ the over-density, $(\rho-\bar\rho)/\bar\rho$.
With just these terms, we recover the standard Poisson equation.

The new terms arising from the non-local distortion function, $\delta\Delta G$, however modify the Poisson equation. To determine these new terms, we see from \ec{DeltaGmn} that we need to evaluate the perturbed $\Delta G_{00}$ or
\bea
\delta\Delta G_{00} &=& \delta\Bigg[ \mathcal{D}_{00}
\Biggl\{ f \Bigr(X\Bigl) + \frac{1}{\square}\Bigl[R f'\Bigr(X\Bigl)\Bigr]\Biggr\}
\vs
+&&\Bigl[ \delta_{0}^{(\rho}\delta_{0}^{\sigma)} 
- \frac{1}{2}g_{00}g^{\rho\sigma}\Bigl] 
\partial_{\rho}\Bigr(X\Bigl)
\partial_{\sigma}\bigg(\frac{1}{\square}\Bigl[R f'\Bigr(X\Bigl)\Bigr]\bigg)\;
\Bigg]\eql{deltag}\vs
\eea
where
\be
\mathcal{D}_{\mu\nu}
\equiv G_{\mu\nu} +g_{\mu\nu}\square - D_\mu D_\nu
.\ee
Capturing all of the relevant first order terms is, for the most part, standard but quite tedious. The inverse d'Alembertian's however introduce a bit of a twist, so we will evaluate the first term as an illustration and then present the final equation.

The first term is
\be
\delta\Delta G_{00}\bigg\vert_{\rm first\,term} 
=
\delta\left(\mathcal{D}_{00} f\right) = \overline{\mathcal{D}}_{00} \delta f + \delta\mathcal{D}_{00} f\big(\overline{X}\big)
.\ee
The simplest piece of this is the one depicted in Fig.~\ref{barx}, $f(\overline{X})$, which -- using \ec{defx} and \ec{deff} -- is a given function of time. The zero order derivative operator is also straightforward; plugging in the FRW metric leads to
\be
\overline{\mathcal{D}}_{00} = 3H^2 + 3H\partial_{t} -\frac{\nabla^2}{a^2} \rightarrow \frac{k^2}{a^2}
\ee
where the first two terms have been dropped because they are smaller than the third by a factor of $(H/k)^2$ (the sub-horizon limit again).
Slightly more complicated is the first order perturbed derivative operator 
\bea
\delta \mathcal{D}_{00}  &=& 6H\dot{\Phi} + 3\dot{\Phi}\partial_{t} 
+ 2\frac{k^2}{a^2}\Phi + 2(\Phi \!-\! \Psi)\frac{\nabla^2}{a^2}
-i\Phi\frac{k_k\partial_k}{a^2}
\vs &\rightarrow &
 2\frac{k^2}{a^2}\Phi
\eea
where the sub-horizon limit kills the first two terms, and the two terms containing spatial derivatives have been set to zero because they vanish when operating on the homogeneous $f(\overline{X})$.
We therefore have
\be
\delta\Delta G_{00}\bigg\vert_{\rm first\,term} 
= \frac{k^2}{a^2} f'(\overline{X}) \delta X 
+ 2\frac{k^2}{a^2}\Phi f\big(\overline{X}\big)
\ee
where we've set $\delta f=f'\times\delta X$.

To obtain $\delta X$, we expand both the d'Alembertian and the Ricci scalar, so that
\bea
\square^{-1} R &=& \osq^{-1} \ovr + \osq^{-1} \delta R -\osq^{-1}\, (\delta\square) \overline{X}.
\eea
The last two terms on the right constitute $\delta X$, so that
\be
\delta X = \osq^{-1} \left[ \delta R - (\delta\square) \overline{X} \right]
.\eql{dx}\ee
Both $\delta R$ and $\delta\square$ can be written in terms of the scalar perturbations. We find in the sub-horizon limit
\bea
\delta R &=&  -2\frac{\nabla^2}{a^2}(\Psi + 2\Phi) \vs
\delta\square &= &0 .
\eql{deltas}\eea
To evaluate $\osq^{-1} \delta R$, it is useful to introduce the retarded Green's function that satisfies
\begin{equation}
\osq G_{\rm ret}(x;x') = 
\Bigl(-\partial_t^2 - 3H \partial_t +\frac{\nabla^2}{a^2} \Bigr)G_{\rm ret}(x;x') 
= \delta^4(x-x')\;.
\end{equation} 
Once we solve for the Green's function, $\delta X$ 
%%%%%%%%%%%%%%%%%%%%%%%%%%%%%%%%%%%
in the sub-horizon limit 
%%%%%%%%%%%%%%%%%%%%%%%%%%%%%%%%%%%
can be written as 
\be
\delta X(x) =  \int\,d^4x'\,G_{\rm ret}(x,x')\, \delta R(x').
\eql{grsol}
\ee

In general the Green's function can be constructed using the massless, minimally coupled scalar mode functions $u(t,k)$ for arbitrary $a(t)$: 
\bea
G_{\rm ret}(x;x') &=& \frac{\Theta(t-t')a^3(t')}{i} \,\int \frac{d^3k}{(2\pi)^3}
e^{i\vec{k}\cdot (\vec{x} - \vec{x'})}\vs
&\times&
\Bigr[ u(t,k)u^*(t',k) - u^*(t,k)u(t',k) \Bigl] \;.
\eea
Although there is no general solution for $G_{\rm ret}$, the Green's function can be captured analytically in the small scale limit (in Fourier space, $k/H_0\rightarrow\infty$)\footnote{In the limit that $\osq^{-1}$ acts on a spatially homogeneous field (in Fourier space, $k/H\rightarrow 0$), the Green's function is $\delta^3(\vec x - \vec x') \int_{t'}^t dt'' \frac{a^3(t'')}{a^3(t')}$. Plugging this expression into the equivalent of \ec{grsol} for $\ovr$ and changing the order of integration leads directly to \ec{defx}.}. Fortunately, this is precisely the limit we are interested in, since it is these scales that can be most easily probed by observations. In this limit, using the WKB approximation we find
\begin{equation}
u(t,k) = \frac{1}{\sqrt{2k}}
\frac{\exp\Bigl[-ik\int^{t}\frac{dt'}{a(t')}\Bigr]}{a(t)} \;.
\end{equation}
so that the integral over $d^4x'$ in \ec{grsol} leads to an expression in Fourier space for the perturbation to $X$:
\be
\delta X(\vec k,t) = \int_{0}^{t} dt' G(t,t';k)\,\delta R(\vec k,t') 
\ee
where the Fourier space Green's function is
\be
G(t,t';k) =
-\frac{1}{k a(t)}a^2(t')
\sin \Bigl[k\int_{t'}^{t}\frac{dt''}{a(t'')}\Bigr]
.\ee
Using \ec{deltas} leads to
\be
\delta X(\vec k,t) = 2\,k^2\int_{0}^{t} \frac{dt'}{a^2(t')} G(t,t';k)
 \left[\Psi(\vec k,t') + 2\Phi(\vec k,t')\right]\eql{dxfin}
.\ee

Treating the other terms in \ec{deltag} similarly leads to our final expression for the modified Poisson equation in the sub-horizon limit:
\be k^2\Phi + k^2E[\Phi, \Psi] 
= 4\pi G a^2 \bar{\rho} \delta.
\eql{modpos}
\ee
where the new terms are functionals of $\Psi$ and $\Phi$:
\bea
E[\Phi,\Psi] &\equiv& 
\Phi \Biggl\{f(\overline{X}) +
\frac{1}{\overline{\square}}\bigg[\overline{R}f'\Bigl(\overline{X}\Bigr)\bigg] \Biggr\}
\vs
&+& \frac{1}{2} % a^2 -> \frac{1}{2}
\Biggl\{
f'(\overline{X})\delta X + 
\osq^{-1} \left[ f'\Bigl(\overline{X}\Bigr)\delta R \right]
 \Biggr\}\eql{defe}
\eea
Note that all the new terms are of order $k^2\Phi$ since the inverse d'Alembertian scales as $k^{-2}$ when acting on the spatially varying $\delta R$ but as $H^{-2}$ when acting on the homogeneous $\ovr f'(\overline X)$.

\subsection{Gravitational Slip Equation}

The equation for the so-called {\it gravitational slip}~\cite{Daniel:2008et,Daniel:2009kr,Bertschinger:2011kk}, the difference between the amplitudes of the two scalar potentials follows from the space-space components of Einstein's (now modified) equations. To derive the modified equation for the gravitational slip, we start with the perturbed space-space components of Einstein's equation, \Ec{ij_delta_eq}. Contracting with $\hat k^i\hat k^j -(1/3)\delta^{ij}$ to extract the longitudinal traceless component leads to
\be
\frac{2}{3a^2}k^2\,
\left( \Phi + \Psi\right) +  
\left[ \hat k^i\hat k^j -(1/3)\delta^{ij}\right] \, \delta\Delta G_{ij} = 8\pi G\left(\bar\rho + \bar P\right) \sigma
\eql{slip}
\ee
where $\sigma$ (in the notation of Ref.~\cite{Ma:1995ey}) represents the anisotropic stress. At late times, when contributions to the energy budget from relativistic species such as photons and neutrinos are negligible, the anisotropic stress vanishes, so $\Phi+\Psi=0$ is a robust prediction from general relativity. 

To quantify the extent to which this prediction is violated in the nonlocal model, we must compute the new terms in \ec{slip}, those proportional to $\delta\Delta G_{ij}$. For the most part, evaluating these requires the same set of techniques used in arriving at the modified Poisson equation. Some of the new ingredients (as always in the sub-horizon limit) are:
\begin{eqnarray}
\overline{\mathcal{D}}_{ij} &=& 
%- sub-horizon limit drop \delta_{ij}a^2(2\dot{H} + 3H^2 +2H\partial_{t}-\partial_{t}^2) 
%+ \delta_{ij}\nabla^2 -\partial_{i}\partial_{j} \vs
-\delta_{ij}k^2 + k_{i}k_{j} \vs 
\delta \mathcal{D}_{ij} &\!\!\!\!=\!\!\!\!& 
- \delta_{ij} k^2(\Phi + \Psi)+ k_{i}k_{j}(\Phi + \Psi).
\end{eqnarray}
Dropping the anisotropic stress, we arrive at the modified equation for the gravitational slip:
\begin{eqnarray}
\lefteqn{(\Phi + \Psi) = 
- (\Phi + \Psi)\Biggl\{f(\overline{X}) +
\frac{1}{\overline{\square}}\bigg[\overline{R}f'\Bigl(\overline{X}\Bigr)\bigg]\Biggl\}}
\nonumber \\
& & 
- f'(\overline{X})\delta X - 
\frac{1}{\overline{\square}}
\Biggl[f'\Bigl(\overline{X}\Bigr)\delta R 
 + \overline{R}f''\Bigl(\overline{X}\Bigr)\delta X \Biggr].
% small scale limit drop: - \delta \square \frac{1}{\overline{\square}}\bigg[\overline{R}f'\Bigl(\overline{X}\Bigr)\bigg]
\eql{anisotropy}
\end{eqnarray}

\section{Solution of the Poisson Equation}\label{sec5}

To solve the modified Poisson equation, note that the new terms due to the non-local interactions are small. This follows from the amplitude of $f$ as depicted in Fig.~\ref{barx} (note that the plotted curve is $10\times f$). We can therefore solve the equation perturbatively by inserting the standard solution for $\Phi$ and $\Psi$ into the new terms and determining their impact on the Poisson equation. This approximation works as long as the effect of the new terms remains small. 

We can write \ec{modpos} as
\begin{equation}
k^2 \Phi \equiv 4\pi G_{\rm eff} a^2 \bar{\rho} \delta %= G_{eff}\frac{k^2\Phi + E[\Phi]}{G}\;.
\end{equation}
so that the fractional change $G_{\rm eff}/G$ is
\begin{equation}
\frac{G_{\rm eff}}{G} = \frac{1}{1+\frac{E[\Phi]}{\Phi}}.
\end{equation}

\begin{figure}[tbh]
\includegraphics[width=7.5cm]{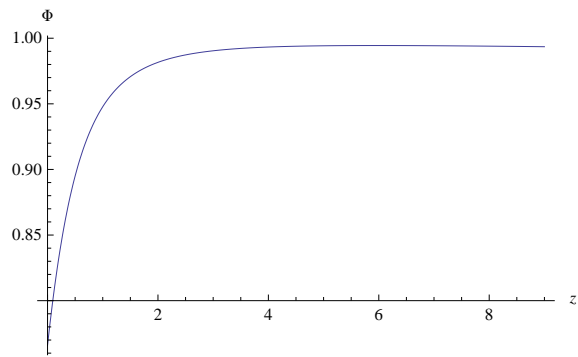}
\caption{The evolution of $\Phi$ is taken from the standard $\Lambda$CDM case.}
\label{Phi}
\end{figure}

The changes to Newton's constant are therefore captured by evaluating the integrals that define $E[\Phi]$ in \ec{defe}. Using \ec{deltas}, \ec{dxfin} and setting $\Psi = -\Phi$ in accordance with our strategy of perturbing around the GR solution leads to
\bea
\frac{E[\Phi]}{\Phi} &=& \left\{f(\overline{X}) +
\frac{1}{\overline{\square}}\bigg[\overline{R}f'\Bigl(\overline{X}\Bigr)\bigg] \right\} 
%+ 2k^2a^2(t) 
+ k^2 \!\int_0^t \frac{dt'}{a^2(t')} 
\vs
&&\times G(t,t';k)\,\frac{\Phi(k,t')}{\Phi(k,t)} \left[ f'(\overline{X}(t))  + f'(\overline{X}(t')) \right].\vs
\eea
To evaluate the integral we use standard linear time dependence of the gravitational potential, as encoded in the growth function:
\be
\Phi(\vec k,a) = \Phi(\vec k,a=1)\,\frac{D_1(a)}{aD_1(a=1)}
\ee
where the growth function in $\Lambda$CDM is
\be
D_1(a) =\frac{5\Omega_mH_0^3}{2} \,\int_0^a\,\frac{da'}{a'^3H(a')^3}
.\ee
%%%%%%%%%%%%Added in v5%%%%%%%%%%%%%%%%%%%%%%%%%%%%%%%%%%%%%%%%%%%%%%%%%%%%
%\textcolor{blue}{
The evolution of $\Phi$ is plotted in Figure~ \ref{Phi}.
%}
%%%%%%%%%%%%%%%%%%%%%%%%%%%%%%%%%%%%%%%%%%%%%%%%%%%%%%%%%%%%%%%%%%%%%%%%%%%

\begin{figure}[tbh]
\includegraphics[width=7.5cm]{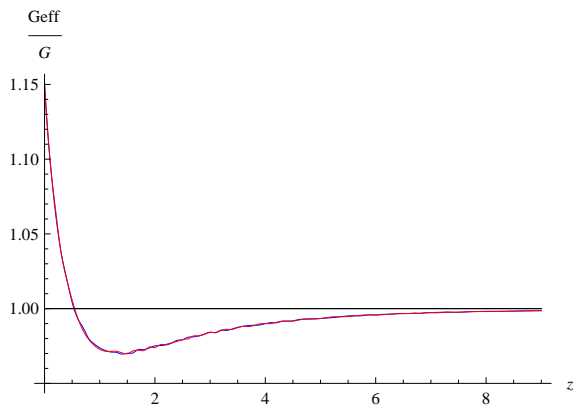}
\caption{The fractional change to Newton's constant, $\frac{G_{\rm eff}}{G}$ as a function of redshift $z$. The two curves, which depict the evolution for $k=0.03, 0.3$h Mpc$^{-1}$, show that the modification is virtually scale-independent, at least in the linear regime.}
\label{Poisson_12_0903}
\end{figure}

In Figure~\ref{Poisson_12_0903} two cases $k=100H_0$ and $k=1000H_0$ are plotted. There is little $k$-dependence, at least on these large scales that are close to linear. The deviations from GR though are quite apparent at late times and should be detectable by upcoming surveys.

%\Sfig{Poisson_12_0903}{The fractional change to Newton's constant, $\frac{G_{\rm eff}}{G}$ as a function of redshift $z$. The two curves, which depict the evolution for $k=0.03, 0.3$h Mpc$^{-1}$, show that the modification is virtually scale-independent, at least in the linear regime.}

%%%%%%%%%%%%Added in v5%%%%%%%%%%%%%%%%%%%%%%%%%%%%%%%%%%%%%%%%%%%%%%%%%%%%%%%
%\textcolor{blue}{
The behavior that $G_{eff}$ drops a little below $G$ for early $z$ can be understood 
by analyzing the effects of terms in the perturbation equations. At zeroth order 
this model has only one mechanism to affect G, which is from the background equation,
\begin{equation}  
\Delta G_{\mu\nu} = \bigg[f(X)+\osq^{-1}(R f'(\overline{X})\bigg]G_{\mu\nu} 
~+~ \mbox{small}
\end{equation}
$G$ is replaced by $\rightarrow~ G/\bigg[1 + f(X)+\osq^{-1}(R f'(\overline{X})\bigg]$
and this is the rescaling of $G$. However at first order there are two effects
in $\delta \Delta G_{\mu\nu}$. The first line of the equation \ec{defe} for $E[\Phi]$  
has the same effect as at the 0th order, which is a rescaling of $G$. 
But the terms on the second line 
%$\frac{k^2}{2}\Biggl\{f'(\overline{X})\osq^{-1}\delta R 
%+ \osq^{-1}f'(\overline{X})\delta R  \Biggr\}$ 
work against the rescaling of $G$.
We can check this by plotting these two kinds of terms. 
Figure~\ref{Poisson_12_0903} is the combination of Figures~\ref{Geff1a2} below. 
%:$\frac{G_{eff}}{G} = 1/[1+f+\osq^{-1}(Rf')+1/2\{f'\osq^{-1}\delta R + \osq^{-1}f'\delta R\}]$
%}
\begin{figure}[tbh]
\includegraphics[width=7.5cm]{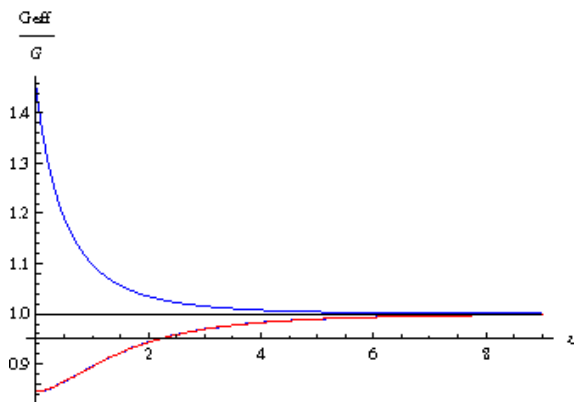}
\caption{Blue curve: $\frac{G_{eff}}{G} = 1/[1+f+\osq^{-1}(Rf')]$ 
considering the first line of \ec{defe}, 
Red and Blue curves below 1: $\frac{G_{eff}}{G} = 1/[1+1/2\{f'\osq^{-1}\delta R + \osq^{-1}f'\delta R\}]$
considering the second line of \ec{defe}.}
\label{Geff1a2}
\end{figure}
%%%%%%%%%%%%%%%%%%%%%%%%%%%%%%%%%%%%%%%%%%%%%%%%%%%%%%%%%%%%%%%%%%%%%%%%%%%%%%%%

\section{Solution for the Gravitational Slip}\label{sec4}

In this section we calculate the gravitational slip $(\Phi +\Psi)$ in the nonlocal model, starting from \ec{anisotropy} and inserting the GR solution for the potentials in all the new terms on the right hand side. The new terms then become
%From the anisotropy equation (\ref{anisotropy}) we identify the four deviated terms as
\begin{eqnarray}
\frac{\Psi+\Phi}{\Phi} &=& 
%-  \frac{2k^2}{\Phi}\int_{0}^{t} \frac {dt'}{a^2(t')} G(t,t';k)\vs
%&&\times
%\Phi(k,t')\left[ f'(\overline{X}(t)) + f'(\overline{X}(t'))\right].
-  \frac{2k^2}{\Phi}\int_{0}^{t} dt'G(t,t';k)\qquad\qquad\qquad\qquad\vs
&\times&
\Biggl\{\frac{\Phi(k,t')}{a^2(t')}
\left[ f'(\overline{X}(t)) + f'(\overline{X}(t'))\right] \Biggr\}.
%\vs
%&+&   \ovr(t')f''(\overline{X}(t'))\!\!\int_{0}^{t'}\!\!dt'G(t',t'';k)\frac{\Phi(k,t'')}{a^2(t'')} 
%\vs
\eea

%\Sfig{aniso_12_0904_noI3}{Gravitational slip as a function of redshift in the nonlocal model. The two curves, barely distinguishable, are for $k=0.03$ h Mpc$^{-1}$ and $k=0.3$ h Mpc$^{-1}$ .}
\begin{figure}[tbh]
\includegraphics[width=7.5cm]{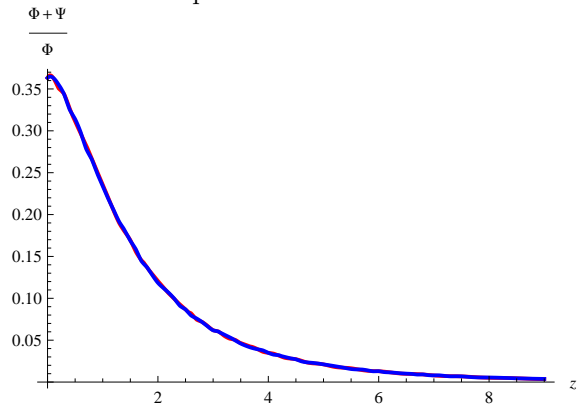}
\caption{Gravitational slip as a function of redshift in the nonlocal model. The two curves, barely distinguishable, are for $k=0.03$ h Mpc$^{-1}$ and $k=0.3$ h Mpc$^{-1}$ .}
\label{aniso_12_0904_noI3}
\end{figure}

Fig.~\ref{aniso_12_0904_noI3} shows the evolution of the gravitational slip as a function of redshift. Here too we find a detectable deviation from the prediction of GR. 

\section{Discussion}

A nonlocal model of gravity is potentially interesting in that it can explain the observed acceleration of the Universe without excessive fine-tuning. In this paper, we have showed that it makes predictions for the growth of structure that differ from those of models based on general relativity. These differences are at the level that the model should be tested by upcoming galaxy surveys.
 
\section*{Acknowledgments}
We are grateful for very helpful discussions with Matthew Dodelson, Mark Trodden, and Richard Woodard.
The computation of the integrals was carried out with the aid of {\it Mathematica}. This work was supported by the U.S. Department of Energy, including grant DEFG02-95ER40896, and by the National Science Foundation under Grant AST-090872. SP was supported by a Fermilab Fellowship in Theoretical Physics.

\bibliography{nonlocal4}

\end{document}